\documentclass[aps,prl,superscriptaddress,footinbib,twocolumn]{revtex4-1} 
\usepackage{graphicx}
\usepackage{amsmath}
\usepackage{braket}
\usepackage{graphicx}
\usepackage{xcolor}  
\usepackage{siunitx}
\usepackage{tabularx}
\usepackage{booktabs}
\usepackage{mathtools}
\usepackage{pifont}

\newcommand{\RN}[1]{%
	\textup{\uppercase\expandafter{\romannumeral#1}}%
}

\makeindex
\begin{document}
\title{Solid-state electron spin lifetime limited by phononic vacuum modes} 
\author{T. Astner}
\email{thomas.astner@tuwien.ac.at}
\affiliation{Vienna Center for Quantum Science and Technology, Atominstitut, TU Wien, Stadionallee 2, 1020 Vienna, Austria}
\author{J. Gugler}
\affiliation{Institute of Applied Physics, TU Wien, Wiedner Hauptstra{\ss}e 8-10/134, 1040 Vienna, Austria}
\author{A. Angerer}
\affiliation{Vienna Center for Quantum Science and Technology, Atominstitut, TU Wien, Stadionallee 2, 1020 Vienna, Austria}
\author{S. Wald}
\affiliation{Vienna Center for Quantum Science and Technology, Atominstitut, TU Wien, Stadionallee 2, 1020 Vienna, Austria}
\author{S. Putz}
\affiliation{Vienna Center for Quantum Science and Technology, Atominstitut, TU Wien, Stadionallee 2, 1020 Vienna, Austria}
\affiliation{Department of Physics, Princeton University, Princeton, NJ 08544, USA}
\author{N. J. Mauser}
\affiliation{Wolfgang Pauli Institute c/o Faculty of Mathematics, Univ. Wien, Oskar Morgensternplatz 1, 1090 Vienna, Austria}
\author{M. Trupke}
\affiliation{Vienna Center for Quantum Science and Technology, Atominstitut, TU Wien, Stadionallee 2, 1020 Vienna, Austria}
\affiliation{Quantum Optics, Quantum Nanophysics \& Quantum Information, Faculty of Physics, Univ. Wien, Boltzmanngasse 5, 1090 Vienna, Austria}
\author{H. Sumiya}
\affiliation{Sumitomo Electric Industries Ltd., Itami, Hyougo, 664-0016, Japan}
\author{S. Onoda}
\affiliation{National Institutes for Quantum and Radiological Science and Technology, 1233 Watanuki, Takasaki, Gunma 370-1292, Japan}
\author{J. Isoya}
\affiliation{Research Centre for Knowledge Communities, University of Tsukuba, 1-2 Kasuga, Tsukuba, Ibaraki 305-8550,
	Japan}
\author{J. Schmiedmayer}
\affiliation{Vienna Center for Quantum Science and Technology, Atominstitut, TU Wien, Stadionallee 2, 1020 Vienna, Austria}
\author{P. Mohn}
\affiliation{Institute of Applied Physics, TU Wien, Wiedner Hauptstra{\ss}e 8-10/134, 1040 Vienna, Austria}
\author{J. Majer}
\email[]{johannes@majer.ch}
\affiliation{Vienna Center for Quantum Science and Technology, Atominstitut c/o Faculty of Mathematics, Univ. Wien, Oskar Morgensternplatz 1, 1090 Vienna, Austria}
\date{\today}

\label{par:intro}
\begin{abstract}
Longitudinal relaxation is the process by which an excited spin ensemble decays into its thermal equilibrium with the environment. In solid-state spin systems relaxation into the phonon bath usually dominates over the coupling to the electromagnetic vacuum. In the quantum limit the spin lifetime is determined by phononic vacuum fluctuations. However, this limit was not observed in previous studies due to thermal phonon contributions or phonon-bottleneck processes. Here we use a dispersive detection scheme based on cavity quantum electrodynamics (cQED) to observe this quantum limit of spin relaxation of the negatively charged nitrogen vacancy ($ \mathrm{NV}^- $) centre in diamond. Diamond possesses high thermal conductivity even at low temperatures, which eliminates phonon-bottleneck processes. We observe exceptionally long longitudinal relaxation times $ T_1 $ of up to \SI{8}{\hour}. To understand the fundamental mechanism of spin-phonon coupling in this system  we develop a theoretical model and calculate the relaxation time \textit{ab initio}. The calculations confirm that the low phononic density of states at the $ \mathrm{NV}^- $ transition frequency enables the spin polarization to survive over macroscopic timescales.

\end{abstract}
\maketitle
The longitudinal relaxation of electron spins trapped in insulating solids has extensively been studied in the past - both theoretically \cite{Waller1932,Elliott1954,Yafet1963,Orbach1961,Overhauser1953} and in experiment \cite{Feher1959,Harrison2006a,Tyryshkin2011,Jarmola2012a,Mrozek2015,Culvahouse1963,Wu2010a,Baral2016}. A theoretical treatment of spin-lattice interaction results in two-phonon Raman \cite{Schuster2007} and Orbach \cite{Orbach1961} processes, dominating at higher temperatures and a direct single phonon process in the low temperature regime. If all thermal phonons are frozen out the ultimate limit of spin lifetime is given by the coupling of the spin system to the phononic vacuum modes. 
The observation of this limit can be obscured by a phonon bottleneck \cite{Scott1962,Tesi2016,Ruby1962}, where excited spins decay and are re-excited if the local phonon excitations are not transferred out of the system sufficiently rapid.

\begin{figure}[t!]
	\includegraphics[width=0.5\textwidth]{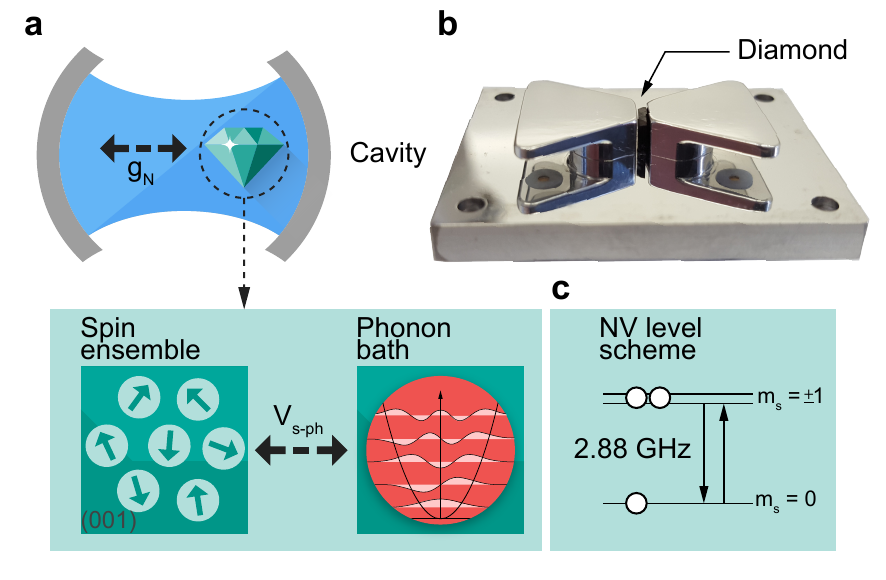}
	\caption{\textbf{Experimental set-up for measuring spin relaxation.} \textbf{a}, The $ \mathrm{NV}^- $ spin ensemble in the diamond crystal exchanges energy with the phonon bath via the spin-phonon interaction potential $ V_{\mathrm{s-ph}} $. The state of the spin ensemble is read out by coupling the spins dispersively to a cavity with a rate $ g_N $. \textbf{b}, Photograph of the superconducting 3D lumped element resonator (top lid removed) with the diamond sample (black). The sample is placed between the metal structures where the magnetic field is homogeneous and all spins couple with the same coupling strength to the resonator mode. \textbf{c}, $ \mathrm{NV}^- $ ground state spin triplet where the $ m_s = \pm 1$ and $ m_s = 0 $ states show a zero field splitting of \SI{2.88}{\giga\hertz}.}
	\label{fig:setup}
\end{figure}

\begin{figure*}[th!]
	\includegraphics[width=0.9\textwidth]{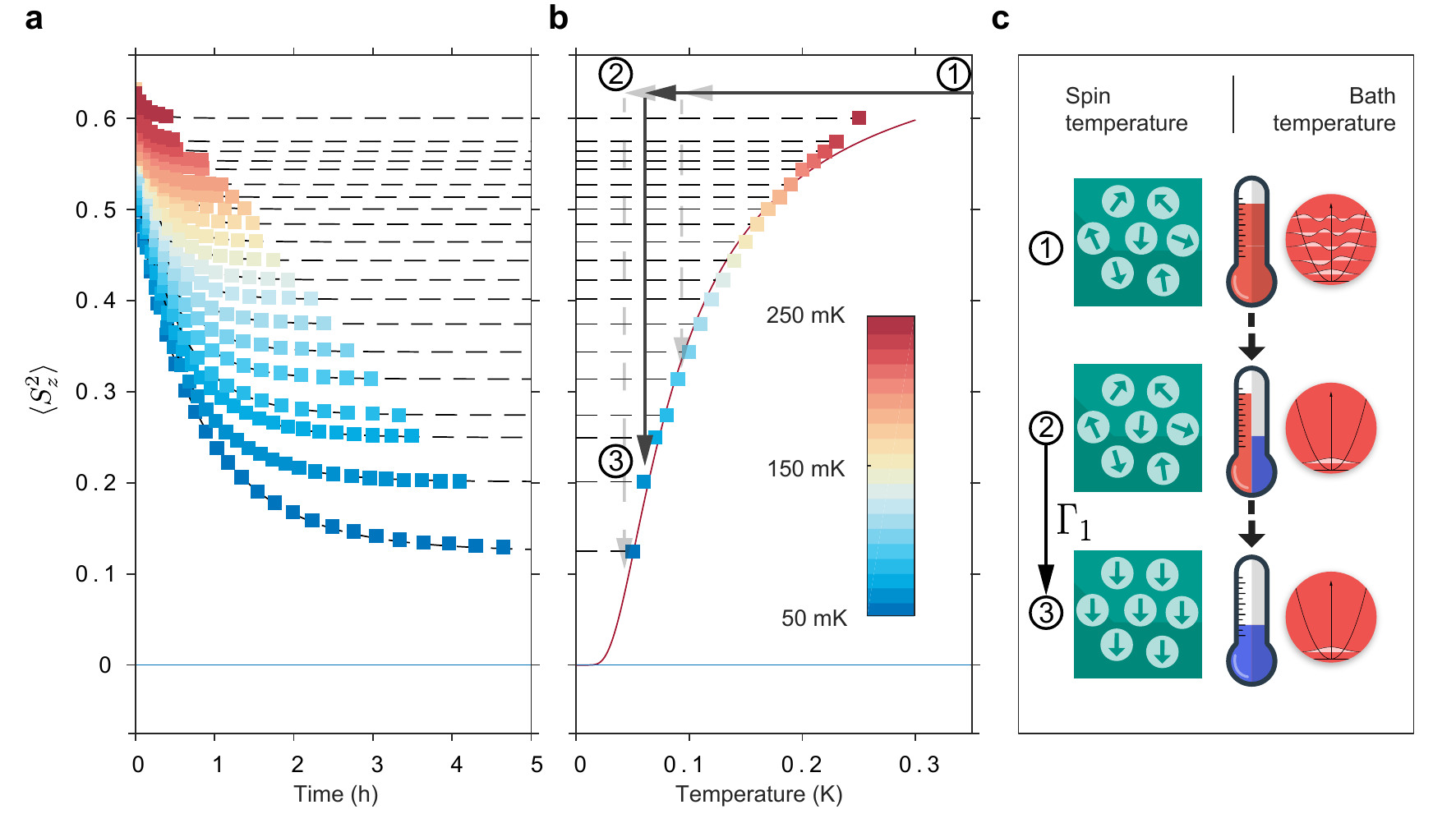}
	\caption{\textbf{Measured time dependence and thermal steady state of $ \braket{S_z^2} $.} \textbf{a}, The time dependent state of the ensemble $ \braket{S_z^2} $ is measured by monitoring the cavity shift. To extract the relaxation rate $ \Gamma $ and the steady state of $ \braket{S_z^2} $ for a given temperature we fit an exponential decay of the form $ \braket{S_z^2(t,T)}-\braket{S_z^2(T)}_{\mathrm{st}} = \left(\braket{S_z^2(T=2.7K)}-\braket{S_z^2(T)}_{\mathrm{st}}\right)e^{-\Gamma t} $. The different colours of the data denote the different target temperatures. \textbf{b-c}, Temperature dependence of the steady state of $ \braket{S_z^2} $: we initialize the system in the steady state at \SI{2.7}{\kelvin} \ding{172}. The phonon bath and the spin ensemble are in thermal equilibrium. Next we non-adiabatically switch to the target temperature \ding{173} to create a cold phonon bath but still excitations in the spin ensemble. The spin system then relaxes into its thermal equilibrium state \ding{174} by exchanging energy with the phonon bath. The red line denotes the steady state solution of $ \braket{S_z^2}_{\mathrm{st}} $ as given in eq.~\ref{eq:steadystate}.}
	\label{fig:inversion}
\end{figure*}

A suitable spin system to observe the coupling to phononic vacuum modes is the $ \mathrm{NV}^- $ centre in diamond. Diamond  has a high thermal conductivity even at low temperatures \cite{Slack1964} which suppresses bottleneck processes. The ground state spin triplet of the $ \mathrm{NV}^- $ centre is split due to dipolar spin-spin interaction by a zero-field splitting of $ \omega_s/2\pi = D/h = \SI{2.88}{\giga\hertz} $ at low temperatures \cite{Ivady2014,Doherty2014} into two upper ($ m_s = \pm 1 $) and a lower ($ m_s = 0 $) level \cite{Jelezko2006b,Ivady2014}. With a base temperature of $ \sim \SI{25}{\milli\kelvin} $, our experimental apparatus can reach temperatures at which the thermal phonon occupation ($ \bar{n} \sim e^{-5.5} \approx \num{4e-4} $) at the spin transition frequency is mostly suppressed ($ \SI{138}{\milli\kelvin} ~\widehat{=}~ \SI{2.88}{\giga\hertz} $).

To investigate the spin-lattice interaction in this system we employ a cQED \cite{Mabuchi2002,Xiang2013a,Kubo2011a,Amsuss2011} protocol where we couple the spins to the oscillating electromagnetic field of a resonator to non-perturbatively read out the spin state as illustrated in Fig.~\ref{fig:setup}a. The interaction of a spin-ensemble with the resonator mode is described by a modified Tavis-Cummings Hamiltonian \cite{Tavis1968} for a three level system with two degenerate excited states (see Methods).
In the dispersive limit the resonator frequency $ \omega_c/2\pi $ and the spin transition frequency $ \omega_s/2\pi $ have a detuning that is much larger than the collective coupling $ g_N $ of the spin ensemble to the resonator mode ($ \delta = \omega_c - \omega_s \gg g_N $). In our experiments we maintain a detuning $ \delta/2\pi > \SI{100}{\mega\hertz} $, corresponding to a negligible enhancement of the single spin spontaneous emission rate due to the Purcell effect \cite{Purcell1946,Bienfait2015} of $ \Gamma_p \approx \SI{e-14}{s^{-1}} $. The dispersive coupling results in a state dependent shift $\chi$ of the resonator frequency given by
\begin{align}
\chi(t,T) &= \frac{N g_0^2}{\delta} (2 - 3\braket{S_z^2(t,T)}).
\label{eq:shift}
\end{align}
The induced shift $\chi$ is enhanced by using a large ensemble of $ N $ spins coupled to the resonator mode with a rate \cite{Dicke1954} $ g_N = g_0 \sqrt{N} $ and a single spin coupling rate of $ g_0 \approx \SI{38}{\milli\hertz} $. The expectation value $ \braket{S_z^2} $ is the measure for the population of spins in the excited $ m_s = \pm 1 $ states.
Therefore, we can continuously monitor the state of the ensemble by probing the cavity resonance frequency \cite{Brune1994,Schuster2007}.
To measure the relaxation time $ T_1 $ we prepare an excited state of the ensemble and monitor the time evolution of the resonator frequency shift $ \chi $ until the ensemble equilibrates with the phonon bath. Fig.~\ref{fig:setup}a illustrates our system where the excited spin ensemble exchanges energy with the phonon bath.

\begin{table*}[th!]
	\caption{\textbf{Sample characteristics.} The samples differ in irradiation treatment, initial nitrogen concentration and final $ \mathrm{NV}^- $ concentration and collective coupling rate $ g_N $ to the resonator mode. The residual nitrogen concentrations $ [\mathrm{N}]_{\mathrm{initial}} - 2[\mathrm{NV}^-] $ are estimated to be \num{120}, \num{20}, \num{24} and \num{180} \si{ppm} for N1, E1, E2, and E3, respectively. The measured rate of spontaneous emission $ \Gamma_0 $ shows a strong dependence on the method of $ \mathrm{NV}^- $ creation. The process of sample preparation is presented in the methods section.}
	\bgroup
	\def\arraystretch{1}
	\setlength\tabcolsep{0.36cm}
	\begin{tabularx}{\textwidth}{lcccc}
		\hline
		\hline
		Sample & N1 & E1 & E2 & E3 \\  
		\hline
		$ \mathrm{NV}^- $[ppm] & 40 & 40 & 13 &  10\\ 
		
		N [ppm] & $ \le $ 200 & 100 & 50 & $ \le $ 200 \\ 
				
		Irradiation type & $n$ & $e^-$& $e^-$&$e^-$\\
		
		$ \Gamma_0 $ [\si{s^{-1}}] & \SI[separate-uncertainty = true,multi-part-units=single]{3.17(10)e-4} & \SI[separate-uncertainty = true,multi-part-units=single]{4.76(26)e-5} & \SI[separate-uncertainty = true,multi-part-units=single]{3.47(16)e-5} &  \SI[separate-uncertainty = true,multi-part-units=single]{7.86(60)e-5}~ \\
		
		$ g_N $ [\si{\mega\hertz}] & \SI[separate-uncertainty = true,multi-part-units=single]{6.62(20)} & \SI[separate-uncertainty = true,multi-part-units=single]{9.11(58)} & \SI[separate-uncertainty = true,multi-part-units=single]{5.88(20)} & \SI[separate-uncertainty = true,multi-part-units=single]{2.85(19)}~ \\ 
		
		Irradiation energy [\si{\mega\electronvolt}]  & \numrange[range-phrase=-]{0.1}{2.5}  & \num{2} & \num{2} & \num{6.5}    \\ 
		
		Annealing temperature [\si{\celsius}]& \SI{900}& \numrange[range-phrase=-]{800}{1000}& \numrange[range-phrase=-]{800}{1000} & \numrange[range-phrase=-]{750}{900}\\
		
		Irradiation dose [\si{cm^{-2}}] & \num{9.0e17}  & \num{1.1e19} & \num{5.6e18} &  \num{1.0e18}\\ 
										
		Mass [\si{mg}]  &  \SI{19.2} & \SI{44.6} & \SI{22.6} & 10.8 \\
		\hline 
		\hline
	\end{tabularx}
	\egroup
	\label{tab:samples}
\end{table*}

The photograph in Fig.~\ref{fig:setup}b shows how the experiment is set-up by bonding a diamond sample into a superconducting 3D lumped element resonator \cite{Angerer2016} fabricated out of aluminium with a fundamental resonance frequency $ \omega_c/2\pi = \SI{3.04}{\giga\hertz} $ and a quality factor $ Q = 60000 $. The resonator geometry focuses the electromagnetic field such that all spins couple homogeneously to the resonator mode. This ensures that spin diffusion effects are negligible in contrast to 2D planar resonator designs \cite{Amsuss2011}. 
The diamond loaded resonator is mounted in an adiabatic demagnetization refrigerator, with temperature regulation between \SI{50}{\milli\kelvin} and \SI{400}{\milli\kelvin} or a dilution refrigerator for temperatures below \SI{50}{\milli\kelvin}.

As presented in Fig.~\ref{fig:inversion}b-c, we measure spin relaxation by first initializing our spin system in a thermal steady state at \SI{2.7}{\kelvin}, a temperature much larger than the transition energy of our system ($ k_B T \gg \hbar\omega_s $). This is equivalent to $ \braket{S_z^2(T\gg \hbar\omega_s/k_B)}_\mathrm{st} \approx 2/3 $. Next, we non-adiabatically switch ($ \tau_{\mathrm{switch}} \ll T_1 $) to a lower target temperature to create a non-equilibrium state of the ensemble (see Methods for details of the temperature regulation).
The time evolution of this state is then given by \cite{Scott1962}
\begin{align}
\frac{\mathrm{d}}{\mathrm{d}t}\braket{S_z^2(t,T)} = -\frac{1}{T_1} \left( \braket{S_z^2(t,T)} - \braket{S_z^2(T)}_{\mathrm{st}} \right),
\label{eq:decay}
\end{align} 
with
\begin{align}
\braket{S_z^2(T)}_{st} &= \frac{2}{e^{\frac{\hbar\omega_s}{k_BT}}+2},
\label{eq:steadystate}
\end{align}
as the temperature dependent steady state of $ \braket{S_z^2(t,T)} $.
We monitor its time evolution via the resonator frequency shift $ \chi $ (see Eq.~(\ref{eq:shift})) until a thermal equilibrium of $ \braket{S_z^2(t,T)} $ is reached. This sequence is repeated for several different target temperatures.

A typical set of experimentally obtained $ \braket{S_z^2(t,T)} $ versus time measurements depicted in Fig.~\ref{fig:inversion}a shows, that an initially prepared non-equilibrium state decays exponentially to its thermal equilibrium value $\braket{S_z^2(t\rightarrow \infty,T)}=\braket{S_z^2(T)}_{st}$ with a characteristic time $ T_1 $.

\begin{figure*}
	\includegraphics[width=1\textwidth]{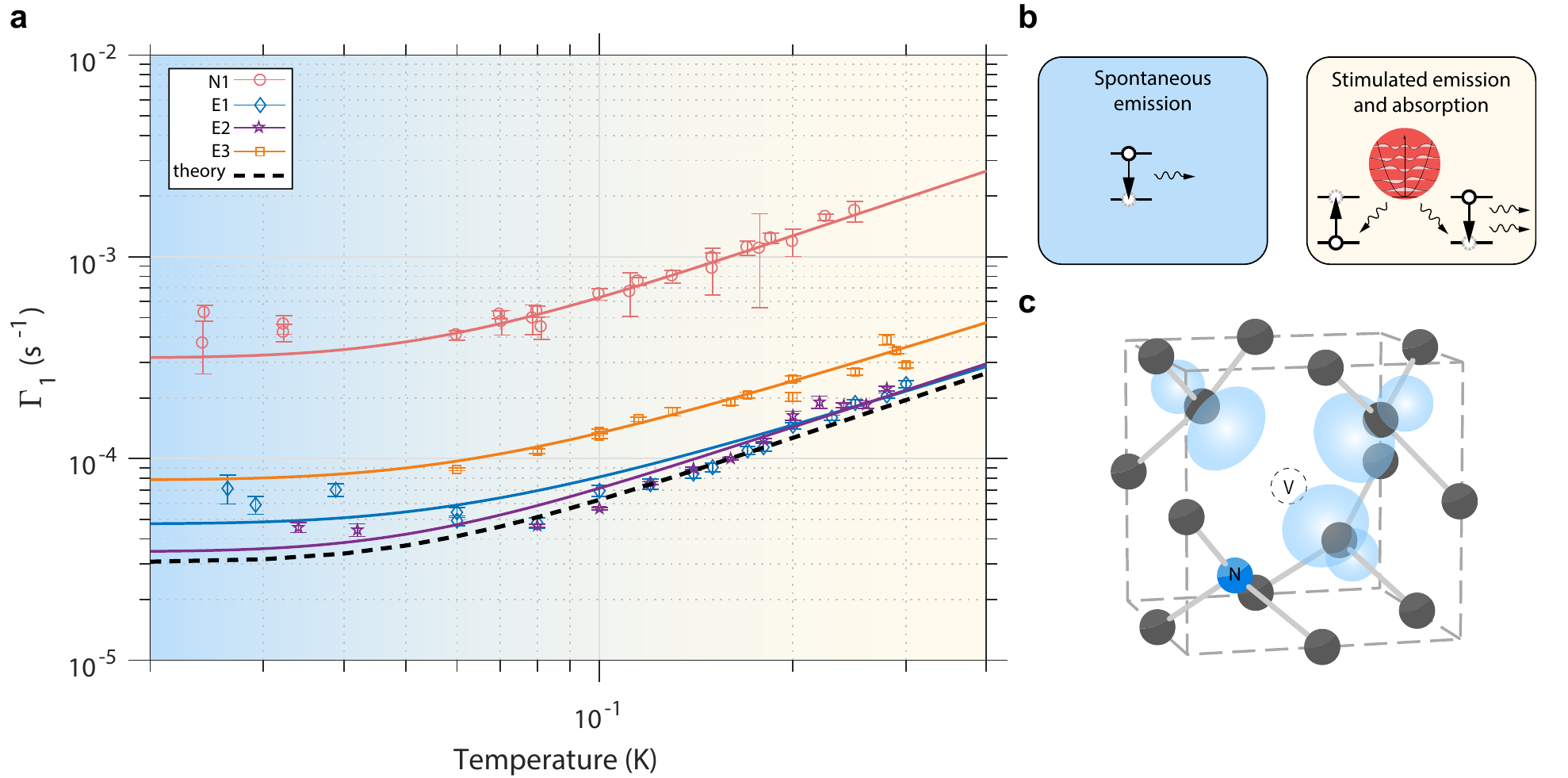}
	\caption{ \textbf{Temperature dependence of the spin-lattice relaxation rate.} \textbf{a}, The symbols represent the measured spin-lattice relaxation rates $ \Gamma $ for different diamond samples. We fit the theoretically predicted temperature dependence $ \Gamma = \Gamma_0 (1+3\bar{n}) $ to extract the factor of proportionality $ \Gamma_0 $. We find the lowest relaxation rate for an electron irradiated sample with $ \Gamma_0 = \num{3.14e-05}$. Note that samples E1 and E2 with different initial nitrogen and $ \mathrm{NV^-} $ concentration exhibit almost the same constant $ \Gamma_0 $, but differ in more than one order of magnitude compared to the neutron irradiated sample N1. The dashed black line corresponds to the relaxation rate calculated \textit{ab intitio}. We identify the energy regime $ k_B T < \hbar \omega_s $ (light blue background) as the quantum limit where the only remaining decay channel is spontaneous emission of a phonon. In the high energy regime $ k_B T > \hbar \omega_s $ (light yellow background) the rate $ \Gamma $ has a linear temperature dependence. This is explained by the high temperature limit of the Bose-Einstein distribution where phonons with an energy of $ \hbar \omega_s $ exist in the phonon bath. In the intermediate regime ($ k_B T \approx \hbar \omega_s $) thermal phonons start to contribute to the relaxation process. \textbf{b}, Illustration of the relevant processes contributing to the spin-lattice relaxation. \textbf{c}, For illustration purposes we show the unit cell containing a single $ \mathrm{NV}^- $ centre with the iso-surface of the spin density shown in blue. The supercell used for the calculation of the theoretical prediction plotted in a, is composed of 64 lattice sites including a single $ \mathrm{NV}^- $ centre.}
	\label{fig:rates}
\end{figure*}

In Fig.~\ref{fig:rates}a we plot the temperature dependence of the relaxation rate $ \Gamma = 1/T_1 $. For temperatures above the spin transition energy $ T > \hbar \omega_s/k_B $ we see a linear dependence of the relaxation rate on temperature. This corresponds to a regime where a direct, single-phonon process dominates. At temperatures $ T < \hbar \omega_s / k_B $ we enter a regime where the relaxation rate is independent of temperature.

To investigate the influence of lattice damage and $ \mathrm{NV}^- $ density on the spin phonon coupling in our system, we perform the same experiment for four different samples. We choose samples that differ in $ \mathrm{NV}^- $ densities and the type of irradiation to create vacancies in the diamond crystal (see Table~\ref{tab:samples}). Samples irradiated by neutrons show much higher lattice damage than electron irradiated ones \cite{Nobauer2013}. In the case of electron irradiation the generated lattice damage depends on the used electron energy and irradiation rate (this can also be seen by comparing sample E3 with E1 and E2 in Table \ref{tab:samples}).
As depicted in Fig.~\ref{fig:rates}a the relaxation rate depends strongly on the lattice damage but is largely independent on the $ \mathrm{NV}^- $ density. This observation indicates that the spin phonon coupling mechanism is intrinsic to the single $ \mathrm{NV}^- $ centre and not a collective effect. Form our experimental data we conclude that the spin-phonon coupling of $ \mathrm{NV}^- $ centres in diamond is weak, resulting in remarkably long spin lifetimes of up to \SI{8}{\hour}.

To understand the underlying mechanism of spin phonon coupling and the measured temperature dependence of $ \Gamma $, we develop a theoretical model and perform \textit{ab initio} calculations to quantify the relaxation rate. We consider the following mechanism \cite{Waller1932} of spin-phonon coupling: The movement of the ions, corresponding to a phononic excitation, affects the positions of the electrons and therefore induces a change of the dipolar spin-spin interaction
\begin{align}
H_{ss}= -\dfrac{\mu_0 g_e^{2} \mu_B^{2}}{4\pi} \dfrac{\left(3(\boldsymbol{S}_{i} \cdot \hat{\boldsymbol{r}}_{ij})(\boldsymbol{S}_{j} \cdot \hat{\boldsymbol{r}}_{ij})-(\boldsymbol{S}_i \cdot \boldsymbol{S}_j)\right)}{\left|\boldsymbol{r}_{ij}\right|^{3}}.
\end{align}
Here $\mu_0$ denotes the permeability of the vacuum, $g_e$ the electronic g-factor, $\mu_B$ the Bohr-magneton, $\boldsymbol{S}_{i(j)}$ is the spin-vector and $\boldsymbol{r}_{ij}(\{\boldsymbol{Q}_n\})$ is the electronic distance vector, which depends on the positions of the ions. Since we are looking at small displacements, the dipolar interaction is expanded to first order with respect to the ionic positions, resulting in a perturbative potential $ V_{\mathrm{s-ph}} $ corresponding to an effective interaction which is then used in Fermis golden rule to calculate transition rates between an initial and a final state $ \Gamma_{f\leftarrow i} $ (see Methods section). The electronic response to the ionic motion is modelled by using the Wigner-Seitz cell as the region around the ionic equilibrium position in which the electronic orbital rigidly follows the ion. Carrying out this procedure, we end up with three relevant processes that describe the dynamics of the spin-lattice interaction in our system: A de-excitation of a spin is accompanied by spontaneous or induced emission of a phonon, whereas the absorption of a phonon excites a spin (see Fig.~\ref{fig:rates}b).
We perform density functional theory (DFT) calculations on a supercell with \num{64} lattice sites and a single $ \mathrm{NV}^- $ centre. The diamond unit cell with a spin density iso-surface is shown in Fig.~\ref{fig:rates}c. In the supercell we calculate the ionic equilibrium positions, the ionic dynamics and the electronic wavefunctions, necessary to quantify $ \Gamma_{f\leftarrow i} $.

Calculating the transition rate gives us a common factor of proportionality $ \Gamma_0 $ for all three processes which also incorporates the phononic density of states.
Spontaneous emission is temperature independent while induced emission and absorption both exhibit a temperature dependence following the Bose-Einstein distribution $\bar{n}$ making them dominant at higher temperatures. Taking all these processes into account we end up with the differential equation (\ref{eq:decay}) and derive an expression for the decay rate $ \Gamma(T) = \Gamma_0 (1+3\bar{n}(T)) $. From our \textit{ab initio} calculations we obtain the value $ \Gamma_0 = \SI{3.02e-5}{s^{-1}} $, in agreement with our measurements for the electron irradiated crystals. These low rates are a consequence of the low phonon density of states at the spin transition energy $\hbar \omega_s $ in a diamond crystal.

Induced emission and absorption govern the linear regime of $\Gamma$ which is explained by the high temperature limit of the Bose-Einstein distribution. The observed plateau at lowest temperatures stems from the temperature independent spontaneous emission of phonons where the decay rate is solely determined by the constant of proportionality $\Gamma_0 $.  Since there are no thermal phonons that match the transition frequency $\omega_s /2\pi$, the measured relaxation rate is inherently limited by quantum fluctuations of the phonon bath.

To conclude, we have shown through the remarkable consistency of experimental data and \textit{ab initio} calculations that in diamond at ultra low temperatures ($ T \ll \hbar \omega_s/k_B $) the longitudinal relaxation time $ T_1 $ of the electron spin is limited by the spontaneous emission of phonons into the phononic vacuum.
Looking at the data in Fig.~\ref{fig:rates} one clearly can extract two main consequences: First, the irradiation and annealing process clearly distinguishes the samples. E1 and E2 (low energy electron irradiation, large dose and high temperature annealing) give $ T_1 $ identical with the \textit{ab intio} calculations. E3 (higher energy electron irradiation with lower fluency) and N (neutron irradiation) show much more lattice damage and shorter $ T_1 $.
Second, it is remarkable that the highest and the lowest $ \mathrm{NV}^- $ density samples have nearly identical curves, indicating that even for high densities the $ T_1 $ processes involve only single spins, and no collective effects.

Even though exceedingly long, $ T_1 $ ($\sim \SI{8}{\hour} $) puts a fundamental limit on the coherence time $ T_2 $ of a spin qubit in diamond. We conjecture that the observed coupling to the phononic vacuum is the dominating process determining $ T_1 $ for spin qubits in magnetically quiet insulators. Tailoring the phononic density of states using phononic meta materials such as phononic band-gaps \cite{Safavi-Naeini2014} and resonators \cite{OConnell2010} may allow to engineer the spin relaxation at ultralow temperatures.

\textbf{~\\Acknowledgements}\\
We thank W. J. Munro, J. Redinger and W. Mayr-Schmoelzer for fruitful discussions. The experimental effort has been supported by the TOP grant of TU Wien and the Japan Society for the Promotion of Science KAKENHI (No. 26246001, 26220903). T.A., A.A. and S.P. acknowledge the support by the Austrian Science Fund (FWF) in the framework of the Doctoral School “Building Solids for Function” (Project W1243). J.G., J.M., N.M. and P.M. acknowledge support by the FWF SFB VICOM (Project F4109-N28). J.S. and N.M. further acknowledge support by the WWTF project “SEQUEX” (Project MA16-066). 



\label{biblio}


\bibliographystyle{mynaturemag}
\bibliography{library}
\newpage
\onecolumngrid
\setcounter{equation}{0}
\newpage
\textbf{~\\Methods and Materials}\\ \\
\textbf{Spin system}\\
The $ \mathrm{NV}^- $ centre in diamond consists of a substitutional nitrogen atom with an adjacent carbon vacancy. In the diamond lattice the $ \mathrm{NV}^- $ centre is oriented along four different crystallographic directions. Its ground state is a paramagentic spin ($ S=1 $) system which we describe with the spin Hamiltonian $ H/h = D S_z^2 $, with a zero field splitting constant $ D = \SI{2.88}{\giga\hertz} $ \cite{Jelezko2006b}. In our experiment we use diamond samples which are cut along the (001) direction and placed in the resonator such that this plane is aligned parallel to the oscillating magnetic field.
\\
\\
\textbf{System Hamiltonian}\\
To describe the system we consider a modified Tavis-Cummings Hamiltonian \cite{Tavis1968} that accounts for the coupling to the degenerate $ m_s = \pm 1 $ spin states of the $ \mathrm{NV}^- $ centre and a term for the cavity probe field with frequency $ \omega_p $ and amplitude $ \eta $:
\begin{align}
	\frac{H_{\mathrm{sys}}}{\hbar} = \underbrace{\omega_\mathrm{c} a^\dagger a + \omega_\mathrm{s} S_\mathrm{z}^2 + i g_N (a^\dagger S^- - a S^+)}_{H_{TC}} + \underbrace{i(\eta a^\dagger e^{-i\omega_p t} - \eta^* a e^{i \omega_p t})}_{H_p},
\end{align}
where $ a/a^\dagger$ are the creation/annihilation operators for the cavity mode with angular frequency $ \omega_c $. 
The operator $ S_\mathrm{z}^2 $ is a collective operator that gives a measure of the spin population in the $ m_s = \pm 1 $ state of an ensemble containing N spins:
\begin{align}
	S_\mathrm{z}^2 = \frac{1}{N} \sum_{j}^{N} \sigma_{z,j}^2.
\end{align}

In the spin basis of the $ \mathrm{NV}^- $ centre ($ \ket{1} $, $ \ket{0} $, $ \ket{-1} $) we define $ \sigma_\mathrm{z}^2 = \ket{1}\bra{1} + \ket{-1}\bra{-1} $, 
and the collective ladder operators $ S^{\pm} $ operators as:
\begin{align}
	S^{\pm} = \frac{1}{N} \sum_{j}^{N} \sigma_j^{\pm}
\end{align}
with $ \sigma^- = \ket{0}\bra{1} + \ket{0}\bra{-1} $ and $ \sigma^+ = \ket{1}\bra{0} + \ket{-1}\bra{0} $.

Thermal excitations of the mode and ensemble are included similarly to \cite{Sandner2012} by adding standard Liouvillian terms to the dynamics and solving the corresponding master equation for the system:
\begin{align}
	\frac{d}{dt} \rho = \frac{1}{i} \left[ H_{\mathrm{sys}},\rho \right] + \mathcal{L}[\rho].
\end{align}
From the master equation we derive the equations of motion for the operators with the detunings $ \Delta_s = \omega_s - \omega_p $ and $ \Delta_c = \omega_c - \omega_p $:
\begin{align}
	\frac{d}{dt}\braket{a} &= -i \Delta_c \braket{a} - i g_N \braket{S^-} + \eta - \kappa \braket{a}, \\
	\frac{d}{dt}\braket{S^-} &=  -i \Delta_s \braket{S^-} - 2i g_N \braket{a} +  3i g_N \braket{a} \braket{S_z^2} - (\gamma_\bot^* + 2 \gamma_\parallel )\braket{S^-}, \\
	\frac{d}{dt}\braket{S_z^2} &=  - i g_N \left( \braket{a} \braket{S^+} - \braket{a^\dagger} \braket{S^-} \right) - 2 \gamma_\parallel \braket{S_z^2} , \\
\end{align}
Next, we calculate the steady-state intra cavity field $ \braket{a}_\mathrm{st} $ for a thermal spin state from which we can extract the spin state dependent cavity shift $ \chi $ in the form:
\begin{align}
\chi(T) &= \frac{N g_0^2}{\delta} (2 - 3\braket{S_z^2(T)}).
\label{eq:shift_methods}
\end{align}
In the equations of motion we use $ \gamma_\bot^* $ for the collective ensemble line width and $ \gamma_\parallel $ for $ T_1 $ processes.
Since the detuning $\delta$ between the spin transition and the cavity is an intrinsic parameter, we can calculate the collective coupling strength $ g_N $ of the spin ensemble to the resonator mode via Eq.~(\ref{eq:shift_methods}).
\\
\\
\textbf{Samples}\\
For the neutron sample N1 we use a commercially available type-Ib high-pressure, high-temperature (HPHT) diamond from the company Element Six Ltd. The crystal contains an initial nitrogen concentration of \SI{<200}{ppm} and a natural abundance of $ ^{13}C $ nuclear isotopes. The local neutron source was a TRIGA Mark II reactor of TU Wien. The sample was irradiated for \SI{50}{\hour} with neutrons in the energy range of \SIrange{0.1}{2.5}{\mega\electronvolt} and a total dose of \num{1.8e18}\si{cm^{-2}}. Next, the sample was annealed for \SI{3}{\hour} at \SI{900}{\degreeCelsius}. The achieved $ \mathrm{NV}^- $ density is \SI{40}{ppm}. In reference \cite{Nobauer2013} (sample BS3-1b) details on the sample preparation are given.
\\
\\
For the electron sample E3 we use a similar raw diamond (HPHT) from element6 as in the case of N1. The crystal contains an initial nitrogen concentration of \SI{<200}{ppm} and a natural abundance of $ ^{13}C $ nuclear isotopes.
We chose an irradiation with \SI{6.5}{\mega\electronvolt} at temperatures of \SIrange{750}{900}{\celsius} with a total fluence of \SI{1e18}{cm^{-2}} which results in a density of \SI{10}{ppm} $ \mathrm{NV}^- $ centre. The irradiation was done at the linear accelerator of the Istituto per la Sintesi Organica e la Fotoreattivita in Bologna, Italy. Details concerning the preparation of this sample can be found in reference \cite{Nobauer2013} (sample U5).
\\
\\
The electron samples E1 and E2 are type-Ib high-pressure, high-temperature (HPHT) diamond crystals with an initial nitrogen concentration of \SI{100}{ppm} and \SI{50}{ppm}, respectively. The sample E1 was irradiated with \SI{2}{\mega\electronvolt} at \SI{800}{\degreeCelsius} and annealed at \SI{1000}{\degreeCelsius} multiple times. For the E2 sample the total electron dose was \SI{1.1e19}{cm^{-2}} and the E1 sample received a total dose of \SI{5.6e18}{cm^{-2}}. The achieved $ \mathrm{NV}^- $ densities are \SI{40}{ppm} and \SI{13}{ppm} respectively.
The electron irradiation was performed by using a Cockcroft-Walton accelerator of the QST, Takasaki. 
\\
\\
\textbf{Microwave resonator}\\
The superconducting 3D lumped element cavity is machined out of aluminium (EN AW 6066) and mechanically polished down to a roughness of $ \approx \SI{0.25}{\micro\meter} $ \cite{Angerer2016}. To bond the crystal samples into the resonator and ensure good thermal contact we use a small amount of vacuum grease as adhesive. We determine the fundamental resonance frequency $ \omega_c/2\pi = \SI{3.04}{\giga\hertz} $ and a quality factor $ Q = 60000 $ by transmission spectroscopy with spin transitions far detuned from the resonator mode.
\\
\\
\textbf{Transmission measurements}\\
In our transmission spectroscopy we determine the forward scattering parameter $ |S_{21}|^2 $ with a standard vector network analyser (R\&S ZNB-8). After cool-down we continuously monitor the transmission to extract the time evolution of the cavity resonance. During the experiment we probe the cavity with an input power of \SI{-110}{dBm} which corresponds to an average number of \SI{1e-9}{} photons/spin in the cavity.
\\
\\
\textbf{Sample initialization procedure}\\
We use an adiabatic demagnetization refrigerator (ADR-Rainier, HPD) with the advantage of temperature regulation in the temperature regime of interest. After measuring the relaxation time for a certain temperature the paramagnetic salt in the fridge has to be magnetized and the fridge is cycled through the \SI{2.7}{\kelvin} heat bath provided by a pulse tube cooler. During this regeneration time of the magnet the spin ensemble is kept at high temperature for initialization. The cooldown time to base temperatures between \SIrange{50}{250}{\milli\kelvin} is in the order of \SIrange{20}{40}{\minute}, which is fast enough that the spins do not completely relax during the cool down process.
Since the ADR is limited to \SI{50}{\milli\kelvin}, the additional data points are measured in a standard dilution refrigerator (Oxford DR-200) where an electric heater allows to thermally initialize the spin ensemble at \SI{1}{\kelvin}. Therefore, we only show a single complete data set of relaxation curves in Fig.~\ref{fig:inversion} which was measured in the adiabatic demagnetization refrigerator. The additional low temperature relaxation measurements in the dilution refrigerator have a different signal amplitude but show the same exponential behaviour.
\\
\\
\textbf{\textit{Ab initio} calculation based on density functional theory}\\
For the calculation of the electron phonon interaction, we consider the change of the electron spin-spin interaction $H_{ss}$ with respect to the ionic movements $\{\boldsymbol{Q}_n\}$.
\begin{align}
H_{ss}(\{\boldsymbol{Q}_n\})=\underbrace{-\dfrac{\mu_0 g_e^{2} \mu_B^{2}}{4\pi}}_{\coloneqq \alpha} \dfrac{\left(3(\boldsymbol{S}_{i} \cdot \hat{\boldsymbol{r}}_{ij}(\{\boldsymbol{Q}_n\}))(\boldsymbol{S}_{j} \cdot \hat{\boldsymbol{r}}_{ij}(\{\boldsymbol{Q}_n\})-(\boldsymbol{S}_i \cdot \boldsymbol{S}_j)\right)}{\left|\boldsymbol{r}_{ij}(\{\boldsymbol{Q}_n\})\right|^{3}}
\end{align}
Here $\mu_0$ is the permeability of the vacuum, $g_e$ the electronic g-factor, $\mu_B$ the Bohr-magneton, $\boldsymbol{S}_{i(j)}$ the spin-vector and $\boldsymbol{r}_{ij}(\{\boldsymbol{Q}_n\})$ is the electronic distance vector dependent on the positions of the ions.
The electronic response to the ionic motion is modelled by defining a region around the ionic equilibrium position $ \boldsymbol{R}_{n}^{(0)}$, in which the electronic orbital rigidly follows the ion. This region is chosen to be the Wigner-Seitz cell. Since we are looking at small displacements, the spin-spin interaction is expanded to first order in the ionic positions. This results in the time dependent perturbative potential $V_{\mathrm{s-ph}}$, which we use in Fermis golden rule to calculate the transition rate $\Gamma_{f\leftarrow i}$:
\begin{align}
\begin{split}
\Gamma_{f\leftarrow i}&=\dfrac{1}{\hbar^{2}}\sum\limits_{f}\lvert \braket{N^{f},m_{s}^{f}|V_{s-ph}|N^{i},m_{s}^{i}}  \rvert^{2} \delta(\nu=\text{\SI{2.88}{\giga\hertz}})\\
&=\dfrac{\alpha^{2}}{\hbar^{2}}\sum\limits_{f}\lvert \bra{m_{s}^{f}}
\sum\limits_{n} \sum\limits_{i>j} \dfrac{ 3(S_{i}^{\pm} S_{j}^{z} + S_{i}^{z} S_{j}^{\pm})(\bra{N^{f}} (Q^{x}_{n} \mp i Q^{y}_{n} ) \ket{N^{i}}) r_{ij}^{z} + (r_{ij}^{x} \mp i r_{ij}^{y} ) \bra{N^{f}} Q^{z}_{n})\ket{N^{i}})\Delta\Theta_{ij}^{n} }{2\left|\boldsymbol{r}_{ij}\right|^{5}}-\\
&-\dfrac{15( S_{i}^{\pm} S_{j}^{z} + S_{i}^{z} S_{j}^{\pm} ) (r_{ij}^{x} \mp i r_{ij}^{y} ) r_{ij}^{z} (\boldsymbol{r}_{ij} \cdot \bra{N^{f}} \boldsymbol{Q}_{n} \ket{N^{i}}) \Delta\Theta_{ij}^{n}}{2\left|\boldsymbol{r}_{ij}\right|^{7}}
\ket{m_{s}^{i}}\rvert^{2} \delta(\nu=\text{\SI{2.88}{\giga\hertz}})\\
\end{split}
\end{align}
Here $m_s^{i(f)}$ and $N^{i(f)}$ denote the initial (final) electronic and phononic states and $\Delta\Theta_{ij}^{n}$ accounts for the fact that the spin-spin interaction between the electronic spins is only changed if either electron $ i $ or electron $ j $ is inside the Wigner-Seitz cell of the displaced ion and the other one outside.
Inserting the second quantized form of the ionic displacement \cite{Mahan2013}
\begin{align}
	\boldsymbol{Q}_{n}=\sum\limits_{\boldsymbol{q},\nu}\sqrt{\dfrac{\hbar}{2M_{n}N\omega_{\boldsymbol{q},\nu}}} \left( a^{\dagger}_{-\boldsymbol{q},\nu}+a_{\boldsymbol{q},\nu} \right) \boldsymbol{\epsilon}_{\boldsymbol{q},\nu}e^{i\boldsymbol{q}\boldsymbol{R}^{(0)}_{n}}
\end{align}
and performing the summation over the final phonon modes leads to a spin-lattice relaxation time $T_{1}$ proportional to the thermal occupation of phonon modes with a frequency of \SI{2.88}{\giga\hertz} for phonon induced emission and absorption and a temperature independent contribution corresponding to spontaneous emission of a phonon with a constant of proportionality $ \Gamma_0 $, dependent on the electronic wavefunctions and the phononic polarization vectors.

The $ \mathrm{NV}^- $ center was simulated \textit{ab initio} in a supercell containing 63 atoms and one vacancy with the Vienna ab initio simulation package (VASP \cite{kresse1996}) using projector augmented wave (PAW) pseudopotentials \cite{bloechl1994} and the exchange-correlation potential of Perdew, Burke and Ernzerhof \cite{perdew1996}. Plane waves up to an energy of \SI{700}{eV} were considered and the Brillouin zone was sampled with a very dense mesh using 19 k-points inside the irreducible wedge. A careful relaxation of the ions to their equilibrium position showed that the nitrogen atoms relax more than the carbon atoms, in agreement with previous calculations \cite {gali2008}. The resulting forces were less than 1 \si{\milli eV/\angstrom} per atom. The results were also confirmed using the Quantum Espresso package \cite{gianozzi2009}.
The diagonalization of the dynamical matrix was done using the PHONOPY-package \cite{togo2015} and the rate $\Gamma_{f\leftarrow i}$ was calculated using maximally localized Wannier functions, obtained with WANNIER90 \cite{mostofi2014}.

\end{document}